\documentclass[floatfix,prl,notitlepage,twocolumn]{revtex4-1}

\usepackage{graphicx}
\usepackage{amsmath}
\usepackage{amsthm}
\usepackage{amssymb}
\usepackage{mathtools}
\usepackage{physics}
\usepackage{bm}
\usepackage{color}
\usepackage{times}
\usepackage[normalem]{ulem}
\usepackage[justification=RaggedRight,font=small]{caption}
\usepackage{subcaption}
\usepackage{tabularx}

\usepackage{appendix}

\usepackage{amstext}
\usepackage{array}

\usepackage{diagbox}

\usepackage{bbm}

\definecolor{wine-stain}{rgb}{0.5,0,0}
\definecolor{bblue}{rgb}{0,0.0,0.5}

\usepackage[colorlinks=true
,urlcolor=bblue
,anchorcolor=blue
,citecolor=wine-stain
,filecolor=blue
,linkcolor=blue
,menucolor=blue
,pagecolor=blue
,linktocpage=true
,pdfproducer=medialab
,linktoc=all 
]{hyperref}

\usepackage{multirow}

\usepackage{booktabs}  

\allowdisplaybreaks

\usepackage{setspace}

\usepackage{custom_commands}

\makeatletter
\def\l@subsubsection#1#2{}
\makeatother

\begin{document}

\title{
Insulator-metal quantum phase transition in heavy topological  insulators
}

\begin{abstract}
Exponentially localized surface states are the most distinctive property of a crystal with non-trivial band topology.
Such surface states play a key role in characterizing topological insulators (TIs), both in theory and experiments.
TIs resulting from the  hybridization of heavy (or nearly flat) and light (or dispersive) bands are automatically tuned to the vicinity of an insulator-to-metal phase transition (IMT), which is \emph{not} accompanied by a change in bulk band-topology.
By formulating a scaling theory for IMTs in such ``heavy" TIs, we show that the proximity to an IMT manifests most dramatically in the behavior of the surface states, viz.  (i) the strength of spin-orbital locking is strongly suppressed; (ii) the surface conduction and valence bands support vastly different number of states; (iii) the surface states penetrate deep into the bulk and the penetration depth diverges at the IMT point. 
Thus, the surface states in heavy TIs coexist with bulk scattered states, and may no longer serve as an useful determinant of their bulk band-topology.  
The mechanism of degradation of surface states discussed here is generic, and expected to hold in both heavy TIs and semimetals.
\end{abstract}

\author{Shouvik Sur$^{1,3}$ and  Pallab Goswami$^{1,2}$}
\affiliation{$^1$Department of Physics \& Astronomy, Northwestern University, Evanston, IL 60208, USA}
\affiliation{$^2$Graduate Program in Applied Physics, Northwestern University, Evanston, IL 60208, USA}
\affiliation{$^3$Department of Physics and Astronomy, Rice University, Houston, Texas 77005, USA}

\date{\today}

\maketitle

One of the striking consequences of a topologically non-trivial,  quasiparticle band-structure is the presence of robust states that are predominantly localized at crystal  terminations  \cite{kane2005, bernevig2006, liu2010a, zhang2012, hasanRev, qiRev, tokuraRev, rachelRev, dzeroRev}.
Traditionally, guided by the bulk-boundary correspondence hypothesis, such surface states have played a crucial role in the detection and characterization of topological materials by both spectroscopic and tunneling  measurements \cite{xia2009, zhang2009, hsieh2009, chen2009, neupane2013, xu2014, pirie2020}, and continues to garner wide interest for their potential for technological applications \cite{xiu2011, gilbert2021}.
In particular, due to their  exponential localization close to surface terminations, surface bound-states can be clearly  distinguished from bulk scattered-states.
Moreover, in topological insulators (TIs) with nearly perfect spectral symmetry between positive and negative energy states, the surface states carry energies that  lie within the bulk band-gap, thereby identifying them as topologically protected mid-gap states.

TIs arising from hybridizing  (nearly) flat and dispersive bands deviate strongly from the spectral-symmetric limit.
Heavy-fermion or Kondo insulators are paradigmatic examples of such TIs, where a nearly dispersionless band, usually resulting from $f$-electrons,  hybridizes with a dispersive, lighter band~\cite{dzero2010,dzero2012,dzeroRev, li2020, paschen2021}.
To what extent do the surface states in such strongly correlated TIs resemble those in their weakly correlated counterparts?
By quantifying the degree of the spectral asymmetry, in this paper, we demonstrate its  non-trivial influence on both surface and bulk properties of such ``heavy" TIs.
Henceforth, we implicitly assume a half-filled system such that the spectral (a)symmetry is equivalent to a particle-hole (a)symmetry about the Fermi energy.

\begin{figure}[!t]
\centering
\includegraphics[width=0.9\columnwidth]{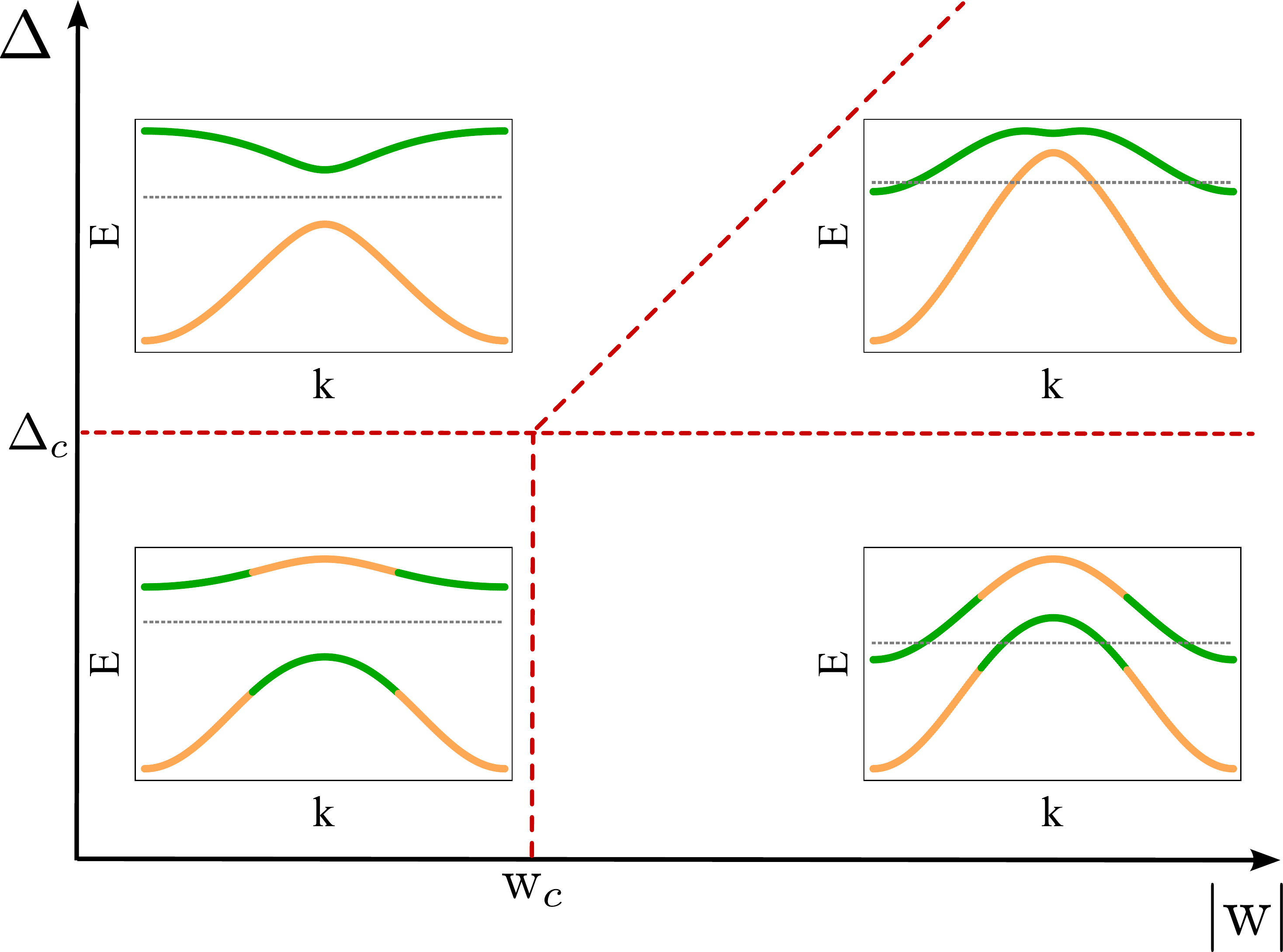}
\caption{Schematic phase diagram of particle-hole asymmetric  insulators.
While $\Delta$ controls band-inversion and tunes a topological phase transition, the asymmetry parameter $\tw$ tunes an  insulator-metal transition (IMT).
At half-filling, four types of phases are possible, viz. topological  and trivial insulators, and topological and trivial compensated metals.
While the topological insulator phase ($\Delta < \Delta_c$ and $|\tw| < \tw_c$) can transition to either the topological metal ($\Delta < \Delta_c$ and $|\tw| > \tw_c$) or  the trivial insulator [$\Delta > \Delta_c$ and $|\tw| < \tw_c(\Delta)$] upon tuning a single parameter, both $\Delta$ and $\tw$ requires tuning to transition to the trivial metal phase [$\Delta > \Delta_c$ and $|\tw| > \tw_c(\Delta)$].
For the class of models represented  in this paper, the transition between the topologically non-trivial (trivial) phases occur at a fixed $\tw_c$ ($\Delta$-dependent $\tw_c$).
In the insets we show representative band-structures in the various regimes of $\Delta$ and $\tw$.
The green and orange colors represent the sign of the  expectation value of the space-inversion operator.
The horizontal dotted lines indicate  the position of the chemical potential  at  half-filling.
}
\label{fig:phase}
\end{figure}

Compared to semimetals~\cite{soluyanov2015, zhang2017}, the manifestation of  particle-hole asymmetries in insulators is less explored.
Here, we develop a theory of a particle-hole asymmetry driven  quantum phase transition between an insulator and a compensated metal in heavy TIs  as depicted in Fig. \ref{fig:phase}. 
We show that proximity to the transition profoundly modifies the behavior of the surface states by simultaneously (i) reducing the strength of spin-orbital locking; (ii) introducing a large imbalance in the number of states constituting the surface conduction and valence bands; (iii) increasing the penetration depth. 
This makes it difficult to both identify the surface state, and distinguish them from bulk scattered states.
Remarkably, sufficiently large particle-hole asymmetries can completely eliminate all   exponentially localized surface states, without changing bulk  band-topology.
Therefore, there exists phases of matter whose band-wavefunctions are topologically equivalent to those in TIs, but do not support gapped or gapless surface states. 
We note that the mechanism for the  suppression of surface states discussed here is generic, and applies to heavy topological insulators and semimetals in all dimensions.

\paragraph*{{\bf Model : } }
Insulators arising from the  hybridization between two spin-degenerate orbitals with distinct effective masses are   described by Hamiltonians of the form  
\begin{align}
H(\bs k) &= \mqty(\veps_1(\bs k) \sig_0 & \vec \eta(\bs k) \cdot \vec \sigma \\ \vec \eta^\dagger(\bs k) \cdot \vec \sigma & \veps_2(\bs k) \sig_0 ),
\end{align}
where $\veps_j(\bs k)$ represents the dispersion of the $j$-th orbital, and $\vec \eta(\bs k)$ encodes the spin and momentum dependent hybridization. 
Here, $\sig_\mu$ acts on the spin degree of freedom with $\sig_j$ ($\sig_0$) being the $j$-th Pauli (identity) matrix.
The Hamiltonian may be  conveniently expressed as
$H(\bs k) = \veps_+(\bs k) \tau_0 \otimes \sigma_0 
+ \veps_-(\bs k) \tau_3 \otimes \sigma_0 + \Re{\vec \eta(\bs k)} \cdot \tau_1 \otimes \vec \sigma + \Im{\vec \eta(\bs k)} \cdot \tau_2 \otimes \vec \sigma$, 
with $\veps_\pm(\bs k) = [\veps_1(\bs k) \pm \veps_2(\bs k)]/2$ and $\tau_\mu$ acting on the orbital degree of freedom.
Assuming the simplest form of the dispersion, $\veps_j(\bs k) =  |\bs k|^2/(2m_j) - \mu_j$, where $m_j$ ($\mu_j$) is the local band-mass (chemical potential) for the $j$-th orbital in the vicinity of a suitable high-symmetry point, we note that   $\bs k$-dependence in the coefficient of the $4\times 4$ identity matrix vanishes only when the two effective masses satisfy $m_1 + m_2 = 0$.
In the limit of interest -- insulators formed by the hybridization of heavy and light bands -- one of the masses dominates over the other, which results in $|(m_1 + m_2)/(m_1 - m_2)| \approx 1$.
Therefore, in heavy TIs the identity matrix is accompanied by a substantial $\bs k$-dependent coefficient.
In the rest of the paper we explore the impact of this term in heavy TIs with $\vec \eta(\bs k) \in \mathbb{R}^3$ .
In particular, we consider a simple form of the hybrization matrix elements, $\eta_j(\bs k) = A_0 k_j$ with $A_0 > 0$, which is suitable for describing TIs with the cubic symmetry.
It is important to note that the heavy-TIs  discussed here are distinguished from previously studied flat-band TIs~\cite{sun2011, ohgushi2000, bergman2008, katsura2010, green2010, bergholtz2013} by the lack of a suppressed bandwidth throughout the Brillouin zone.

In analogy to the $k.p$ models of first order topological insulators, we define $B = (m_2 - m_1)/(4 m_1 m_2)$, $B' = (m_2 + m_1)/(4 m_1 m_2)$, 
$M_0 = (\mu_2 - \mu_1)/2$, and 
$\mu = (\mu_2 + \mu_1)/2$, such that 
\begin{align}
H(\bs k) = (B' |\bs k|^2 - \mu) \mathbbm{1} + (B |\bs k|^2 + M_0) \Gamma_5 + A_0 \sum_{j=1}^3 k_j \Gamma_j,
\end{align}
where the mutually anticommuting matrices $\{\Gamma_{j \leq 3}, \Gamma_5 \} = \{\tau_1 \otimes \sigma_j, \tau_3 \otimes \sigma_0 \}$. 
At half-filling, the ground state of $H(\bs k)$ is topologically non-trivial (trivial) if  $B$ and $M_0$ have opposite (same) sign.
This is succinctly expressed through the sign of the band-inversion parameter $\Delta \coloneqq M_0/B$, which vanishes at the topological quantum critical point separating the two phases \cite{murakami2008, goswami2011}.
Here, we introduce a second parameter that encodes the degree of particle-hole asymmetry,
\begin{align}
\tw \coloneqq B'/B = \frac{1 + m_1/m_2}{1 - m_1/m_2}.
\end{align}
A perfect particle-hole symmetric insulator is obtained only at $\tw = 0 \equiv m_2 + m_1 = 0$.
As $(m_1/m_2) \to 0$ or $\infty$, the system draws progressively closer to  an insulator-to-metal transition (IMT) at $|\tw| = 1$ for any $\Delta < 0$.
For $\tw > 1$ with $\Delta < 0$ the system ceases to be an insulator  irrespective of filling, and is best described as a ``topological metal".
At half-filling it possesses both electron and hole pockets.
Importantly, the  metallic phase is topologically equivalent to the topological insulator, because $\tw$ does not affect the Bloch wavefunctions of the bands.
In the thermodynamic limit, at half-filling, the IMT is characterized by a sharp jump in the density of states at the Fermi level, which changes from $0$ at $|\tw| < 1$ to a finite, non-universal quantity for $|\tw| > 1$.
Therefore, the IMT is a sharply defined phase transition in the bulk.
In Fig.~\ref{fig:phase} we present a schematic phase diagram, and note that in the topologically trivial part of the phase diagram (i.e. $\Delta > 0$) the location of the IMT depends on both $\tw$ and $\Delta$.
Here, we will only focus on the IMT between topologically non-trivial phases ($\Delta < 0$).

\begin{figure}[!t]
\centering
\includegraphics[width=0.9\columnwidth]{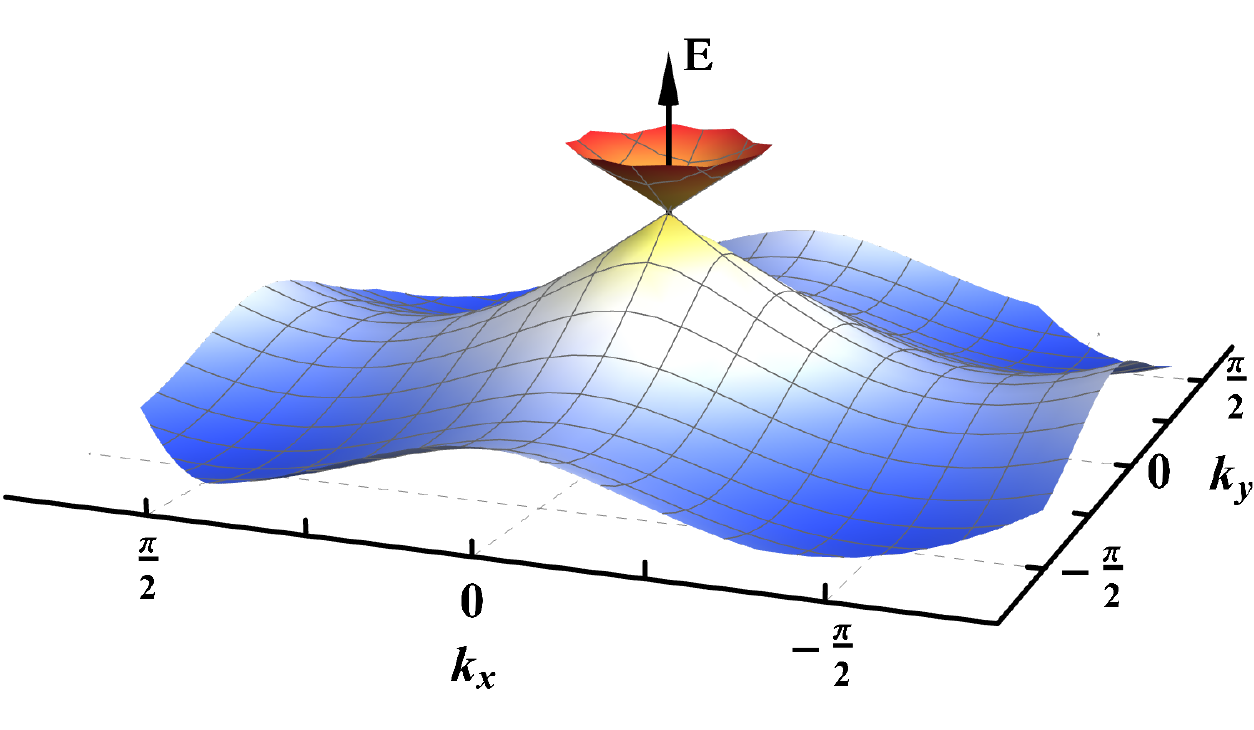}
\caption{Deformation of the surface states due to particle-hole asymmetry.
The most prominent effect of the asymmetry manifest through an inequivalent normalizability conditions on states that constitute the two cones of the Dirac-like surface states of a topological insulator (TI).
For the specific choice of parameters in this paper, the number of states in the upper/conduction (lower/valence) band is suppressed (increased).
}
\label{fig:cone}
\end{figure}


\paragraph*{{\bf Surface states :}}
While the topological properties of the bulk bandstructure are unaffected by the particle-hole asymmetry, the surface states are sensitive to all terms in the Hamiltonian \cite{zhou2008, linder2009}.
The strength of the particle-hole asymmetry strongly influences the dispersion of the surface states, their penetration into the bulk, and the magnitude of spin-orbital locking (SOL). 
In particular, a divergent penetration depth indicates a lack of surface localization.
In the topological insulator phase exponentially localized states exist over a finite sub-region of the surface Brillouin zone (sBZ).
By contrast, the topological metal does not support any exponentially localized state, and exemplifies a topological phase of matter without gapped or gapless surface states.

\begin{figure}[!t]
\centering
\begin{subfigure}[b]{0.8\columnwidth}
\centering
\includegraphics[width=\columnwidth]{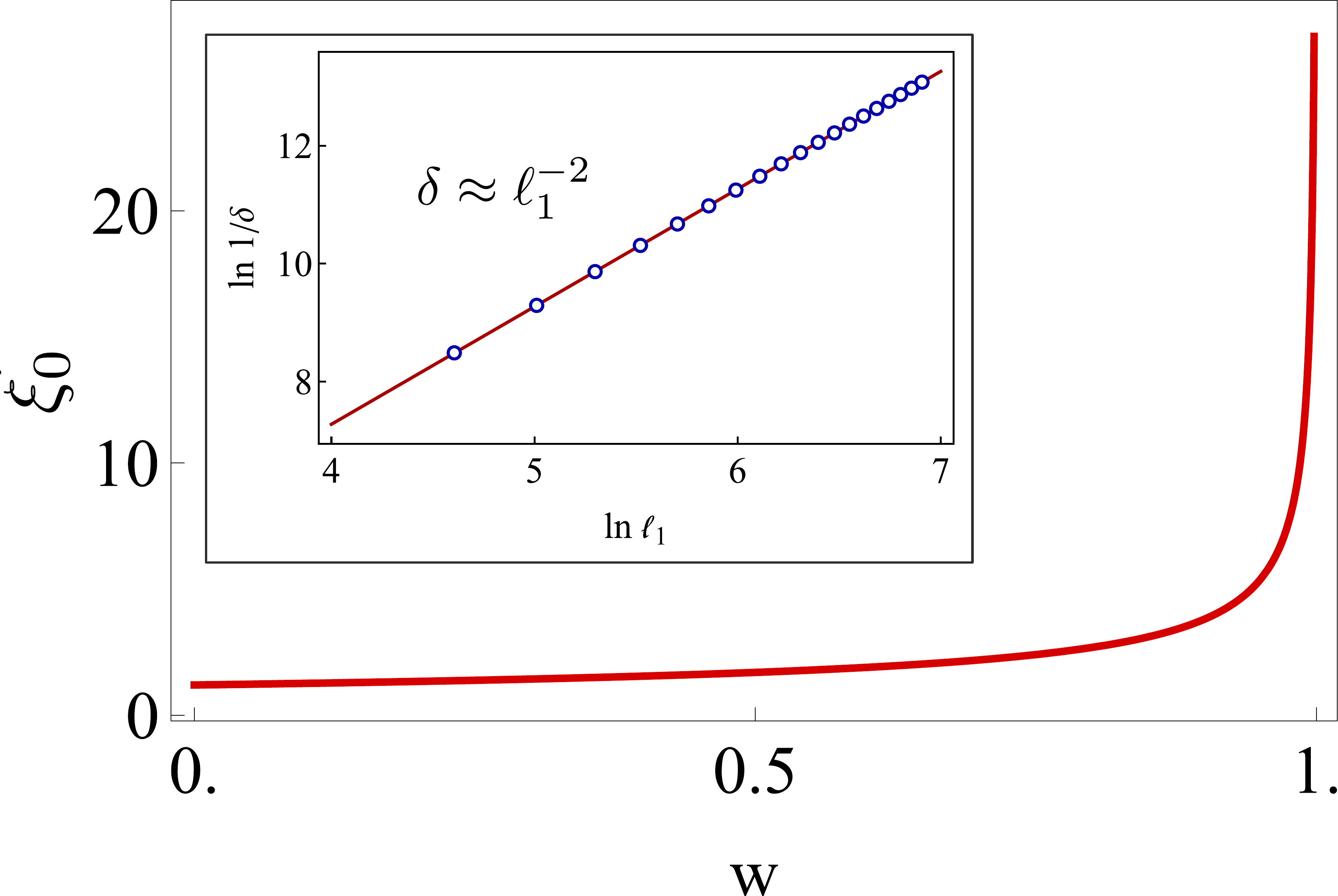}
\caption{}
\label{fig:loc}
\end{subfigure}
\hfill
\begin{subfigure}[b]{0.9\columnwidth}
\centering
\includegraphics[width=\columnwidth]{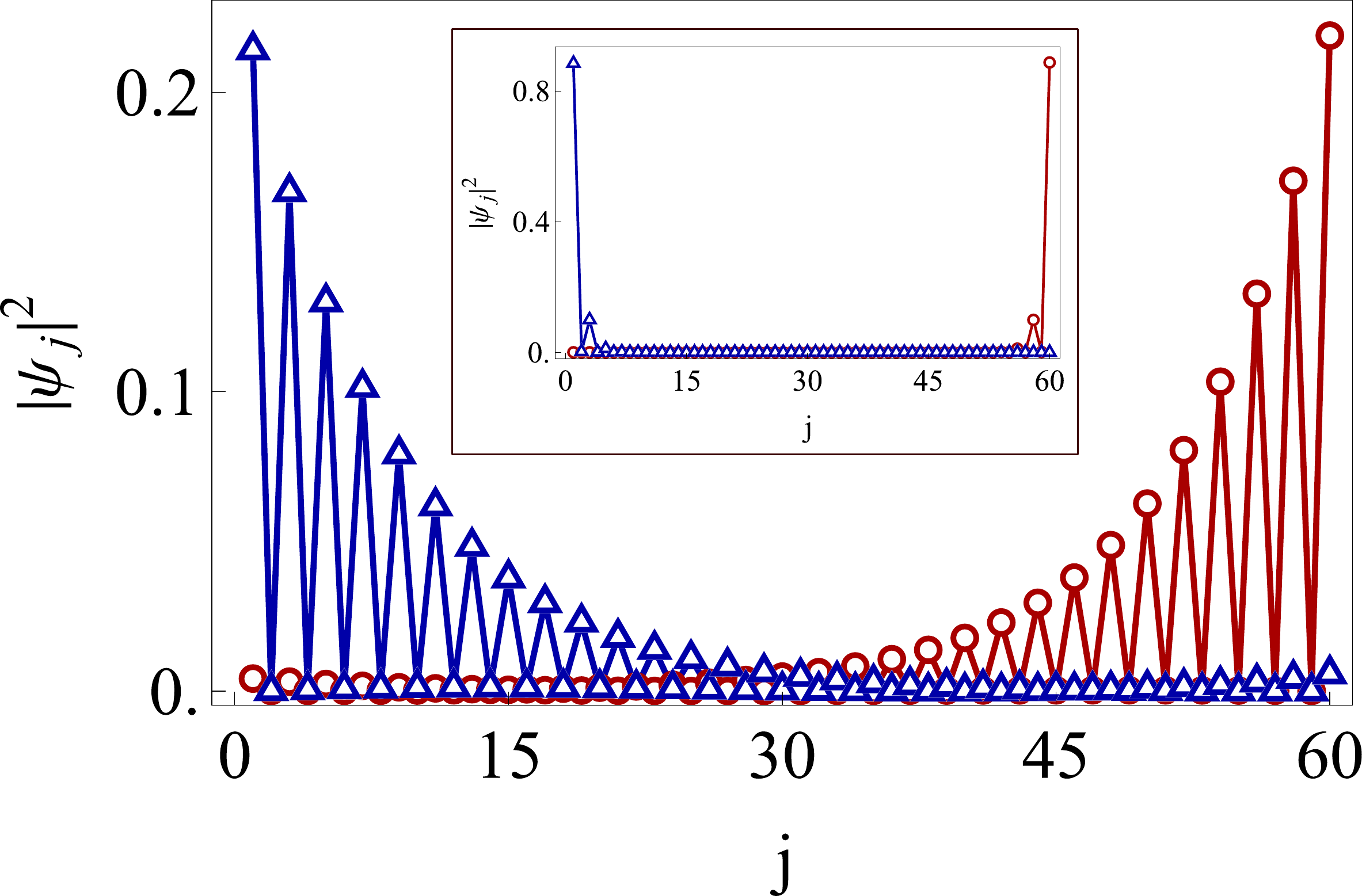}
\caption{}
\label{fig:amp}
\end{subfigure}
\caption{Increase of the penetration depth of the surface states with proximity to the insulator-metal transition, and its implications. 
(a) The penetration depth of the states forming the surface Dirac points diverges as $1/\sqrt{1-\tw}$.  In systems of finite size, $L$, a spectral gap, $\dl$, is present at the location of the surface Dirac points. (inset) In the limit of large system size, $\ell_1 \coloneqq L/a \gg 1$ ($a$ is the unit-cell size), $\dl$ is suppressed by a factor of $\ell_1^{-\alpha}$ with $\alpha > 1$.
Here, we have set $\xi = 0.99 L$ and $\dl$ is measured in units of $2B a^2$.
(b) The probability amplitude of the surface states progressively increases in the bulk-layers. Here, we have numerically obtained the amplitude of the wavefunctions of the surface zero-mode at $\tw = 0.995$ (which corresponds to $\xi(0) \approx 17$) by exact diagonalization in a slab of thickness $L = 60 a$ along the $[100]$ direction.
}
\label{eq:PD}
\end{figure}

In order to analytically demonstrate these results, we consider a semi-infinite TI governed by $H(x_1, \bs k_\perp)$ with the bulk in the $x_1 > 0$ half-space, where $x_1$ is the position-vector component conjugate to  $k_1$.
In order to elucidate the behavior of the surface state as a function of $\tw$, it is convenient to choose a suitable basis in which $H(x_1, \bs k_\perp)$ emulates the Rice-Mele model~\cite{rice1982} at fixed $\bs k_\perp$, 
\begin{align}
H_{\bs k_\perp}'(\partial_1) &= \qty[\mu(\bs k_\perp) -  \tw B \partial_1^2 ] \mathbbm{1} 
+ \qty[M(\bs k_\perp) - B \partial_1^2] \Gamma_5 \nn \\
& - i A_0 \partial_1 \Gamma_1
+  C(\bs k_\perp) \Gamma_{2},
\label{eq:RM}
\end{align}
where $\mu(\bs k_\perp) = \tw B |\bs k_\perp|^2 - \mu_0$, $M(\bs k_\perp) = M_0 + B|\bs k_\perp|^2$, $C(\bs k_\perp) = A_0 |\bs k_\perp|$ with $\bs k_\perp \coloneqq (k_2, k_3)$.
Without loss of generality, we assume $B$ and $\tw$ to be positive real numbers. 
The Rice-Mele insulators governed by Eq.~\eqref{eq:RM} are  continuously parameterized by $\bs k_\perp$, and for a fixed $\bs k_\perp$ the insulator is topological non-trivial only if $\text{sign}(M(\bs k_\perp)) \neq \text{sign}(B)$.
The eigenvalue equation $H_{\bs k_\perp}'(\partial_1) \ket{\Psi(x_1, \bs k_\perp)} = E_{\text{surf}}(\bs k_\perp) \ket{\Psi(x_1, \bs k_\perp)}$ is solved for states that are exponentially localized near $x_1 = 0$ and decay into the bulk. 
These states disperse as
\begin{align}
E_{\text{surf}}^\pm(\bs k_\perp) = \mu(\bs k_\perp) - \tw M(\bs k_\perp) \pm \sqrt{1 - \tw^2} ~C(\bs k_\perp),
\end{align}
and they are restricted to those sBZ wavevectors that satisfy $\Re{\lambda_{1,\pm}(\bs k_\perp)} > 0$, where 
\begin{align}
\lambda_{1,\pm}(\bs k_\perp) = \frac{A_0 - \sqrt{A_0^2 + 4 B f_\pm(\bs k_\perp)}}{2B \sqrt{1-\tw^2}} 
\end{align}
with $f_\pm(\bs k_\perp) =   (1 - \tw^2)M(\bs k_\perp) \pm  \tw \sqrt{1-\tw^2} C(\bs k_\perp)$.
This inequality determines whether an exponentially localized state exists  at a given $\bs k_\perp$.
Consequently, the localization length or penetration depth of the  surface states at $\bs k_\perp$ is given by  
\begin{align}
\xi(\bs k_\perp) = \mbox{max}\qty{\frac{1}{\Re[\lambda_{1,+}(\bs k_\perp)]}, \frac{1}{\Re[\lambda_{1,-}(\bs k_\perp)]} }.
\end{align}

At $\tw = 0$,  a particle-hole symmetric, non-degenerate Dirac cone is obtained, which extends over the region $|\bs k_\perp|^2 < - M_0/B$.
As the topological quantum critical point for the Rice-Mele insulator labeled by $\bs k_\perp$ is approached, $m(\bs k_\perp) \coloneqq \frac{M(\bs k_\perp)}{A_0^2/B} \to 0^-$ and $\xi(\bs k_\perp) \sim - \frac{B}{A_0} m^{-1}(\bs k_\perp)$.
Thus, the ratio $\frac{\xi(\bs k_\perp)  }{L} \to \infty$, where $L$ controls the system size and $L/a \gg 1$ with $a$ being the lattice spacing.
The implications for the 2D surface states of the 3D TI are twofold, viz. (i) for a generic $M_0 < 0$, as the boundary of the region in the sBZ supporting normalizable surface states is approached, the penetration depth  diverges as $(\sqrt{|M_0|/B} - |\bs k_\perp|)^{-1}$; (ii) upon tuning toward the topological critical point at $M_0 = 0$, the states at the surface Dirac point extend deeper into the bulk and remain appreciable over a scale $\sim A_0/|M_0|$.

For $\tw > 0$, the leading impact of  particle-hole asymmetry  on the surface states manifest through  $\tw$ dependence of the magnitude of spin-orbit coupling, spectra, and the normalizability condition, $\Re{\lam_{1,\pm}} > 0$.
In particular, the surface Hamiltonian describing the normalizable states close to the zone-center of the sBZ is given by
\begin{align}
H_{\text{surf}}(\bs k_\perp) &= \qty[\mu(\bs k_\perp) - \tw M(\bs k_\perp)] \rho_0  \nn \\
& \quad + A_0 \sqrt{1 - \tw^2} \qty[k_2 \rho_3 + k_3 \rho_2],
\end{align}
where $\rho_\mu$ acts on the space of surface conduction and valence bands.
Thus, the magnitude of SOL in the surface states is suppressed by a factor of $\sqrt{1 -\tw^2}$.
Analogously, the magnitude of the Fermi velocity decreases as the IMT is approached from the TI side,
\begin{align}
v_F^{\pm}(\bs k_\perp) \coloneqq |\bs \nabla E_{\text{surf}}^\pm(\bs k_\perp)| = A_0 \sqrt{1 - \tw^2} + \ordr{|\bs k_\perp|},
\end{align} 
which leads to a gradual flattening of the Dirac cone as $\tw \to 1^-$.
Thus, the surface modes in nearly flat-band TIs result in heavy Dirac  quasiparticles \cite{alexandrov2013}.
Here, since both $\mu(\bs k_\perp)$ and $\tw M(\bs k_\perp)$ affect the dispersion at quadratic order in $|\bs k_\perp|^2$, they do not lead to an asymmetry of the Dirac cone at the leading order.
Instead, the dominant influence of the particle-hole asymmetry is encoded in the normalizability condition, which has opposite effects on the number of states in the conduction and valence bands of the Dirac cone, as exemplified by Fig. \ref{fig:cone}.
In particular, the upper (lower) cone supports fewer (more)  normalizable states than the $\tw = 0$ limit, because $f_+$ ($f_-$) is negative-valued over a smaller (larger) range of $\bs k_\perp$.
Consequently, surface states exist at a fixed $|\bs k_\perp| < |M_0|/B$ as long as $|\tw| < \tw_c(\bs k_\perp)$, where  $\tw_c(\bs k_\perp) = \sqrt{1 - \frac{C^2(\bs k_\perp)}{M^2(\bs k_\perp) + C^2(\bs k_\perp)}} $.
This implies, the Dirac band-crossing at $\bs k_\perp = 0$ survives only for $\tw < \tw_c(0) = 1$.
Since $\tw_c(\bs k_\perp) \leq 1$,  $\tw_c(0)$ marks the asymmetry strength beyond which \emph{no surface state can exist}.
The penetration depth of the gapped surface states at a fixed $|\bs k_\perp| > 0$ diverges as $\xi(\bs k_\perp) \sim \frac{A_0 \chi(\bs k)}{M^2(\bs k_\perp) + C^2(\bs k_\perp)} [\tw_c(\bs k_\perp) - \tw]^{-1}$ as $\tw$ approaches $\tw_c(\bs k_\perp)$ from below.
Thus, the states at the Dirac point decay into the bulk over a scale \begin{align}
\xi_0 \coloneqq \xi(0) =  - \frac{A_0/M_0}{\sqrt{2(1-\tw)}} + \ordr{\sqrt{1-\tw}}.
\end{align} 
We depict its behavior in Fig. \ref{fig:loc}.
As a result of the gentler decay, the probability of detecting surface states deep in the bulk increases with  proximity to the IMT, as shown in Fig. \ref{fig:amp}.

\paragraph*{{\bf Critical fan and finite-size effects :}} One of the striking consequences of quantum criticality is the presence of a critical fan (CF) that emanates from the critical point. 
Here, the IMT point at $\tw = 1$ seeds an CF, which dictates the crossover behaviors on the TI side of the phase diagram.
The energy scale for the CF is set by $\xi_0^{-1}$, and it vanishes as $\tw \to 1$~\cite{note1}.
Thus, at a fixed distance from the topological quantum critical point, three distinct length scales are generically present in particle-hole asymmetric TIs, viz. unit cell size ($a$), system size ($L$), and $\xi_0$.
Here, we assess the finite size effects in terms of $\ell_1 \coloneqq L/a$ and $\ell_2 \coloneqq \xi_0/L$, and consider their manifestation through the gapping of the surface Dirac points due to hybridization between surface states originating from opposite ends of a sample in the slab geometry.
In the thin-film limit $\ell_1 \sim 1$, and previous findings continue to hold~\cite{zhou2008, linder2009,liu2010b, lu2010, shan2010}, independent of $\ell_2$.
By contrast, in the thermodynamic limit $\ell_1 \gg 1$, and the spectral gap at the location of the surface Dirac point $\delta \sim \frac{A_0^2}{B}  \ell_1^{-\alpha} \exp{-b \ell_2^{-\beta}}$ with $\alpha, \beta \gtrsim 1$ and $b$ a positive number.
The information about the CF is encoded in the exponential factor: deep in the TI phase [CF] $\ell_2 \ll 1$ [$\ell_2 \gtrsim 1$] and $\exp{-b \ell_2^{-\beta}} \to 0$ [an $\ordr{1}$ number].
Interestingly, because of the $\ell_1$-dependent suppression factor, in the thermodynamic limit $\dl \to 0$ with $\ell_1$ at any  fixed value of $\ell_2$ (see the inset of Fig. \ref{fig:loc}).
Therefore, for a sufficiently large system with $\ell_2 \sim 1$, $\dl$ can effectively vanish even though the  surface states penetrate deep inside the bulk. 

\begin{figure}[!t]
\centering
\includegraphics[width=0.67\columnwidth]{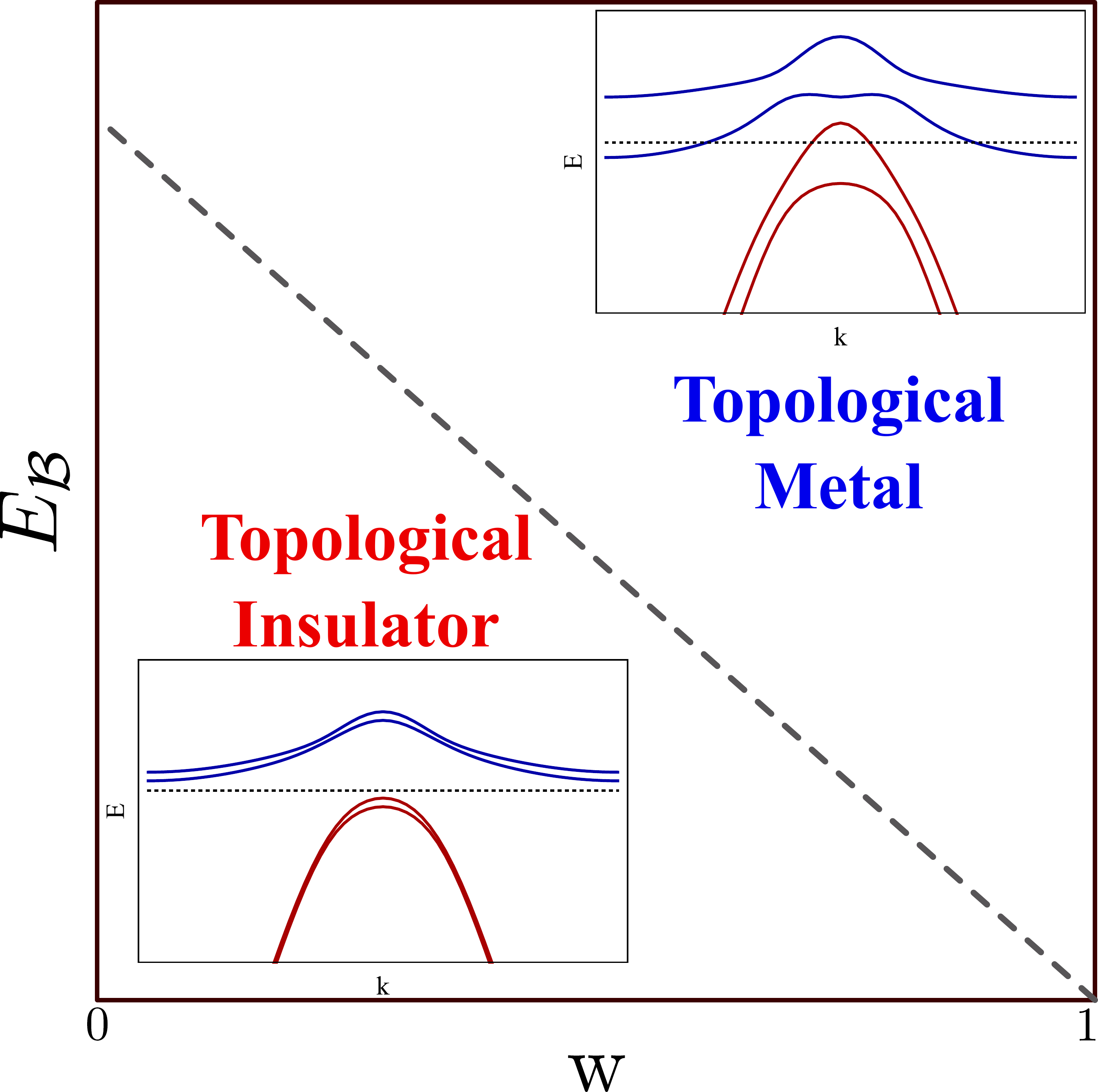}
\caption{Phase diagram in the presence of an externally applied magnetic field. 
The magnetic field spin-splits the bands, which decreases the energy gap between the conduction and valence sectors, and pushes the insulator towards a metallic state.
The minimum strength of the magnetic field necessary (dashed line) to induce the corresponding insulator-metal transition is proportional to $(1-\tw)$.
(Inset) Representative band-structure in the insulating and metallic phases; horizontal dotted lines indicates the location of the chemical potential at half-filling. 
}
\label{fig:BvsW}
\end{figure}


\paragraph*{{\bf Discussion :}}
Owing to the essential participation of heavy orbitals in the formation of heavy-fermion topological insulators and semimetals~\cite{xu2017, lai2018, guo2018}, these classes of compounds are naturally tuned to the vicinity of an IMT or an IMT-like transition.
In such situations the properties of conventional TIs and semimetals may prove to be qualitatively insufficient for describing the emergent physics.
Here, we demonstrated an important aspect of such revisions necessary to correctly describe the surface physics of heavy TIs.
In particular, the surface states  penetrate deep into the bulk without opening a hybridization gap at the surface Dirac points in thermodynamically large systems, and, thus, spatially coexist with the bulk scattered states.
Since the surface states still exist as mid-gap states in the TI phase, their response would appear to originate from both the bulk of a sample and the mid-gap region of the spectrum. 
It is, therefore, conceivable that many heavy-fermion insulators would exhibit apparently anomalous bulk response resulting from  deeply buried surface states, and, thus, our results are potentially relevant to observations reported in Refs. \cite{park2016, laurita2016}.
Combined with the suppressed spin-orbital locking strength, our results also imply that surface-states-based diagnoses of bulk topology in heavy TIs require  more care.
For example, a few cleavage surfaces are experimentally accessible, and they may lack  surface states due to a particle-hole asymmetry driven breakdown of  normalizability, which is not equivalent to an absence of topological protection.

We note that Zeeman-type spin splitting of energy levels provides an additional path toward an IMT in the presence of an external magnetic field. 
The critical strength of the magnetic field necessary to drive the transition decreases with $(1-\tw)$, such that TIs with large particle-hole asymmetries ($\tw \approx 1^-$) are particularly susceptible to a magnetic-field driven IMT.
In Fig.~\ref{fig:BvsW} we elucidate this cooperation between the  particle-hole asymmetry and externally applied magnetic field in  inducing an IMT in heavy TIs.
Furthermore, in analogy to the surface states discussed above, particle-hole asymmetries facilitate deeper penetration of the surface Landau levels into the bulk.
In particular, the end-modes supported by the zeroth Landau level can be obtained analytically, and we find that its penetration depth in the limit $\tw \to 1^{-}$ scales as 
$\xi_{\mc B} \sim (1-\tw)^{-1/2}$.

The scope of our results extend beyond three dimensional heavy TIs, and applies with minor modifications to both three dimensional heavy semimetals, as well as lower dimensional topological systems. 
In particular, the chiral edge states of a hypothetical heavy chern insulator would develop a strongly asymmetric distribution of states at positive vs. negative values of edge momenta, in addition to penetrating deep into the bulk. 
By extrapolation, the chiral Fermi arcs in heavy, time-reversal symmetry broken Weyl semimetals would inherit the anomalous behaviors of the edge states of the embedded heavy Chern insulating planes lying between the Weyl points, and potentially support unusual phenomenology.
A similar set of anomalies is expected to arise in heavy higher order topological matter, particularly affecting hinge and corner localized states.
An exploration of such possibilities is left for future works. 

\paragraph*{{\bf Acknowledgments :}}
This work was supported by the National Science Foundation MRSEC program (DMR-1720139) at the Materials Research Center of Northwestern University. The work of S.S. at Rice University was supported by the U.S. Department of Energy Computational Materials Sciences (CMS) program under Award Number DE-SC0020177. A part of this work was performed at the Aspen Center for Physics, which is supported by the National Science Foundation grant PHY-1607611.


%
%


\begin{thebibliography}{99}
\bibitem{hasanRev} M. Z. Hasan and C. L. Kane, Rev. Mod. Phys. {\bf 82}, 3045 (2010).
\bibitem{qiRev} X.-L. Qi and S.-C.  Zhang, Rev. Mod. Phys. {\bf 83}, 1057  (2011).
\bibitem{tokuraRev} Yoshinori Tokura, Kenji Yasuda, and Atsushi Tsukazaki, Nat.  Rev. Phys. {\bf 1}, 126 (2019).
\bibitem{rachelRev} S. Rachel, Rep. Prog. Phys. {\bf 81}, 116501 (2018).
\bibitem{dzeroRev} M. Dzero, J. Xia, V.  Galitski, and P. Coleman, Annu. Rev. Condens. Matter Phys. {\bf 7}, 249 (2016).
\bibitem{kane2005} C. L. Kane and E. Mele, Phys. Rev. Lett.  {\bf 95}, 226801 (2005).
\bibitem{bernevig2006} B. A. Bernevig and S.-C. Zhang, Phys. Rev. Lett. {\bf 96}, 106802 (2006).
\bibitem{liu2010a} C.-X. Liu, X.-L. Qi, H. Zhang, X. Dai, A. Fang, and S.-C. Zhang, Phys. Rev. B {\bf 82}, 045122 (2010).
\bibitem{zhang2012} F. Zhang, C. L. Kane, E. J. Mele, Phys. Rev. B {\bf 86}, 081303(R) (2012).
\bibitem{xia2009} Y. Xia, D. Qian, D. Hsieh, L. Wray, A. Pal, H. Lin, A.
Bansil, D. Grauer, Y. S. Hor, R. J. Cava, and M. Z. Hasan, Nat. Phys. {\bf 5}, 398 (2009).
\bibitem{zhang2009} H. Zhang, C.-X. Liu, X.-L. Qi, X. Dai, Z. Fang, and
S.-C. Zhang, Nat. Phys. {\bf 5}, 438 (2009).
\bibitem{hsieh2009} D. Hsieh, Y. Xia, D. Qian, L. Wray, J. H. Dil, F. Meier, J. Osterwalder, L. Patthey, J. G. Checkelsky, N. P. Ong, A. V.
Fedorov, H. Lin, A. Bansil, D. Grauer, Y. S. Hor, R. J. Cava,
and M. Z. Hasan, Nature (London) 
{\bf 460}, 1101 (2009).
\bibitem{chen2009} Y. L. Chen, J. G. Analytis, J.-H. Chu, Z. K. Liu, S.-K. Mo, X. L. Qi, H. J. Zhang, D. H. Lu, X. Dai, Z. Fang, S. C.
Zhang, I. R. Fisher, Z. Hussain, and Z.-X. Shen, Science {\bf 325}, 178 (2009).
\bibitem{neupane2013} M. Neupane, N. Alidoust, S-Y. Xu, T. Kondo, Y. Ishida, D. J. Kim, Chang Liu, I. Belopolski, Y. J. Jo, T-R. Chang, H-T. Jeng, T. Durakiewicz, L. Balicas, H. Lin, A. Bansil, S. Shin, Z. Fisk, and M. Z. Hasan, Nat.  Commun.  {\bf 4}, 2991 (2013).
\bibitem{xu2014} N. Xu, P. K. Biswas, J. H. Dil, R. S. Dhaka, G. Landolt, S. Muff, C. E. Matt, X. Shi, N. C. Plumb, M. Radovi\'{c}, E. Pomjakushina, K. Conder, A. Amato, S. V. Borisenko, R. Yu, H.-M. Weng, Z. Fang, X. Dai, J. Mesot, H. Ding, and M. Shi, Nat.  Commun. {\bf 5}, 4566 (2014).
\bibitem{pirie2020} H. Pirie, Y. Liu, A. Soumyanarayanan, Pengcheng Chen, Yang He, M. M. Yee, P. F. S. Rosa, J. D. Thompson, Dae-Jeong Kim, Z. Fisk, Xiangfeng Wang, Johnpierre Paglione, Dirk K. Morr, M. H. Hamidian, and  Jennifer E. Hoffman, Nat. Phys. {\bf 16}, 52 (2020).
\bibitem{xiu2011} F. Xiu, L. He, Y.  Wang, L. Cheng, L.-T. Chang, M. Lang, G. Huang, X. Kou, Y. Zhou, X. Jiang, Z. Chen, J. Zou, A. Shailos, and K. L. Wang, Nat. Nanotechnol. {\bf 6}, 216 (2011).
\bibitem{gilbert2021} M. J. Gilbert, Commun. Phys.  {\bf 4}, 70 (2021).
\bibitem{dzero2010} M. Dzero, K. Sun, V.  Galitski, and P. Coleman, Phys. Rev. Lett. {\bf 104}, 106408 (2010).
\bibitem{dzero2012} M. Dzero, K. Sun, P. Coleman, and V. Galitski, Phys. Rev. B {\bf 85}, 045130 (2012).
\bibitem{li2020} L. Li, K. Sun, C. Kurdak, and J. W. Allen, Nat. Rev. Phys. {\bf 2}, 463 (2021). 
\bibitem{paschen2021} S. Paschen and  Q. Si, Nat. Rev. Phys. {\bf 3}, 9 (2021).
\bibitem{soluyanov2015} A. A. Soluyanov, D. Gresch, Z. Wang, Q. Wu, M. Troyer, X. Dai, B. A. Bernevig, Nature (London) {\bf 527}, 495-498 (2015).
\bibitem{zhang2017} K. Zhang, G.E. Volovik, J. Low Temp. Phys. {\bf 189}, 276 (2017).
\bibitem{ohgushi2000} K. Ohgushi, S. Murakami, and N. Nagaosa
, Phys. Rev. B {\bf 62}, R6065(R) (2000).
\bibitem{bergman2008} D. L. Bergman, C. Wu, and L. Balents
, Phys. Rev. B {\bf 78}, 125104 (2008).
\bibitem{katsura2010} H. Katsura, I.  Maruyama, A. Tanaka, and H. Tasaki, Europhys Lett. {\bf 91}, 57007 (2010).
\bibitem{green2010} D. Green, L.  Santos, and C. Chamon, Phys. Rev. B {\bf 82}, 075104 (2010).
\bibitem{sun2011} K. Sun, Z.-C. Gu, H. Katsura, S. Das Sarma, Phys. Rev. Lett. {\bf 106}, 236803 (2011).
\bibitem{bergholtz2013} E. J. Bergholtz, and Z. Liu, Int. J. Mod. Phys. B {\bf 27}, 1330017 (2013).
\bibitem{murakami2008} S. Murakami and S. Kuga, Phys. Rev. B {\bf 78}, 165313 (2008).
\bibitem{goswami2011} P. Goswami and S. Chakravarty, Phys. Rev. Lett. {\bf 107}, 196803 (2011).
\bibitem{zhou2008} B. Zhou, H.-Z. Lu, R.-L. Chu, S.-Q. Shen, and Q. Niu, Phys. Rev. Lett. {\bf 101}, 246807 (2008)
\bibitem{linder2009} J. Linder, T. Yokoyama, and A. Sudb\o{}, Phys. Rev. B {\bf 80}, 205401 (2009).
\bibitem{rice1982} M. J. Rice and E. J. Mele, Phys. Rev. Lett. {\bf 49}, 1455 (1982).
\bibitem{liu2010b} C.-X. Liu, H.  Zhang, B. Yan, X.-L. Qi, T. Frauenheim, X. Dai, Z. Fang, and S.-C. Zhang, Phys. Rev. B {\bf 81}, 041307(R) (2010).
\bibitem{lu2010} H.-Z. Lu, W.-Y.  Shan, W. Yao, Q. Niu, and S.-Q. Shen, Phys. Rev. B {\bf 81}, 115407 (2010).
\bibitem{shan2010} W.-Y. Shan, H.-Z. Lu, S.-Q. Shen, New J. Phys. {\bf 12}, 043048 (2010).
\bibitem{alexandrov2013} V.  Alexandrov, M. Dzero, and P. Coleman, Phys. Rev. Lett. {\bf 111}, 226403, (2013)
\bibitem{note1} A similar energy scale also arises due to the topological quantum critical point, and it is encoded in the band-inversion parameter $\Delta$.
\bibitem{xu2017} Y. Xu, C. Yue, H. Weng, and X. Dai, Phys. Rev. X {\bf 7}, 011027 (2017).
\bibitem{lai2018} H.-H. Lai, S. E. Grefe, S. Paschen, and Q. Si, Proc. Natl. Acad. Sci. {\bf 115}, 93 (2018).
\bibitem{guo2018} C. Y. Guo, F. Wu, Z. Z. Wu, M. Smidman, C. Cao, A. Bostwick, C. Jozwiak, E. Rotenberg, Y. Liu, F. Steglich, and H. Q. Yuan, Nat. Comm. {\bf 9}, 4622 (2018).
\bibitem{park2016} W. K. Park, L.  Sun, A. Noddings, D.-J. Kim, Z. Fisk, and L. H. Greene, Proc. Natl. Acad. Sci. {\bf 113},  6599 (2016).
\bibitem{laurita2016} N. J. Laurita, C. M. Morris, S. M. Koohpayeh, P. F. S. Rosa, W. A. Phelan, Z. Fisk, T. M. McQueen, and N. P. Armitage, Phys. Rev. B {\bf 94}, 165154 (2016).
\end{thebibliography}
\end{document}